\documentclass[aps,prd,amsmath,amssymb,reprint,showpacs,superscriptaddress,nobibnotes,floatfix]{revtex4-1}

\usepackage{graphicx}
\usepackage{dcolumn}
\usepackage{bm}
\usepackage{hyperref}
\usepackage[mathlines]{lineno}
\usepackage{siunitx}

\usepackage{scrextend}

\begin{document}

\preprint{APS/123-QED}

\title{Measurement of the inclusive-isolated prompt-photon cross section in \boldmath{$ p\bar{p} $} collisions \\ 
using the full CDF data set }




\affiliation{Institute of Physics, Academia Sinica, Taipei, Taiwan 11529, Republic of China}
\affiliation{Argonne National Laboratory, Argonne, Illinois 60439, USA}
\affiliation{University of Athens, 157 71 Athens, Greece}
\affiliation{Institut de Fisica d'Altes Energies, ICREA, Universitat Autonoma de Barcelona, E-08193, Bellaterra (Barcelona), Spain}
\affiliation{Baylor University, Waco, Texas 76798, USA}
\affiliation{Istituto Nazionale di Fisica Nucleare Bologna, \ensuremath{^{kk}}University of Bologna, I-40127 Bologna, Italy}
\affiliation{University of California, Davis, Davis, California 95616, USA}
\affiliation{University of California, Los Angeles, Los Angeles, California 90024, USA}
\affiliation{Instituto de Fisica de Cantabria, CSIC-University of Cantabria, 39005 Santander, Spain}
\affiliation{Carnegie Mellon University, Pittsburgh, Pennsylvania 15213, USA}
\affiliation{Enrico Fermi Institute, University of Chicago, Chicago, Illinois 60637, USA}
\affiliation{Comenius University, 842 48 Bratislava, Slovakia; Institute of Experimental Physics, 040 01 Kosice, Slovakia}
\affiliation{Joint Institute for Nuclear Research, RU-141980 Dubna, Russia}
\affiliation{Duke University, Durham, North Carolina 27708, USA}
\affiliation{Fermi National Accelerator Laboratory, Batavia, Illinois 60510, USA}
\affiliation{University of Florida, Gainesville, Florida 32611, USA}
\affiliation{Laboratori Nazionali di Frascati, Istituto Nazionale di Fisica Nucleare, I-00044 Frascati, Italy}
\affiliation{University of Geneva, CH-1211 Geneva 4, Switzerland}
\affiliation{Glasgow University, Glasgow G12 8QQ, United Kingdom}
\affiliation{Harvard University, Cambridge, Massachusetts 02138, USA}
\affiliation{Division of High Energy Physics, Department of Physics, University of Helsinki, FIN-00014, Helsinki, Finland; Helsinki Institute of Physics, FIN-00014, Helsinki, Finland}
\affiliation{University of Illinois, Urbana, Illinois 61801, USA}
\affiliation{The Johns Hopkins University, Baltimore, Maryland 21218, USA}
\affiliation{Institut f\"{u}r Experimentelle Kernphysik, Karlsruhe Institute of Technology, D-76131 Karlsruhe, Germany}
\affiliation{Center for High Energy Physics: Kyungpook National University, Daegu 702-701, Korea; Seoul National University, Seoul 151-742, Korea; Sungkyunkwan University, Suwon 440-746, Korea; Korea Institute of Science and Technology Information, Daejeon 305-806, Korea; Chonnam National University, Gwangju 500-757, Korea; Chonbuk National University, Jeonju 561-756, Korea; Ewha Womans University, Seoul, 120-750, Korea}
\affiliation{Ernest Orlando Lawrence Berkeley National Laboratory, Berkeley, California 94720, USA}
\affiliation{University of Liverpool, Liverpool L69 7ZE, United Kingdom}
\affiliation{University College London, London WC1E 6BT, United Kingdom}
\affiliation{Centro de Investigaciones Energeticas Medioambientales y Tecnologicas, E-28040 Madrid, Spain}
\affiliation{Massachusetts Institute of Technology, Cambridge, Massachusetts 02139, USA}
\affiliation{University of Michigan, Ann Arbor, Michigan 48109, USA}
\affiliation{Michigan State University, East Lansing, Michigan 48824, USA}
\affiliation{Institution for Theoretical and Experimental Physics, ITEP, Moscow 117259, Russia}
\affiliation{University of New Mexico, Albuquerque, New Mexico 87131, USA}
\affiliation{The Ohio State University, Columbus, Ohio 43210, USA}
\affiliation{Okayama University, Okayama 700-8530, Japan}
\affiliation{Osaka City University, Osaka 558-8585, Japan}
\affiliation{University of Oxford, Oxford OX1 3RH, United Kingdom}
\affiliation{Istituto Nazionale di Fisica Nucleare, Sezione di Padova, \ensuremath{^{ll}}University of Padova, I-35131 Padova, Italy}
\affiliation{University of Pennsylvania, Philadelphia, Pennsylvania 19104, USA}
\affiliation{Istituto Nazionale di Fisica Nucleare Pisa, \ensuremath{^{mm}}University of Pisa, \ensuremath{^{nn}}University of Siena, \ensuremath{^{oo}}Scuola Normale Superiore, I-56127 Pisa, Italy, \ensuremath{^{pp}}INFN Pavia, I-27100 Pavia, Italy, \ensuremath{^{qq}}University of Pavia, I-27100 Pavia, Italy}
\affiliation{University of Pittsburgh, Pittsburgh, Pennsylvania 15260, USA}
\affiliation{Purdue University, West Lafayette, Indiana 47907, USA}
\affiliation{University of Rochester, Rochester, New York 14627, USA}
\affiliation{The Rockefeller University, New York, New York 10065, USA}
\affiliation{Istituto Nazionale di Fisica Nucleare, Sezione di Roma 1, \ensuremath{^{rr}}Sapienza Universit\`{a} di Roma, I-00185 Roma, Italy}
\affiliation{Mitchell Institute for Fundamental Physics and Astronomy, Texas A\&M University, College Station, Texas 77843, USA}
\affiliation{Istituto Nazionale di Fisica Nucleare Trieste, \ensuremath{^{ss}}Gruppo Collegato di Udine, \ensuremath{^{tt}}University of Udine, I-33100 Udine, Italy, \ensuremath{^{uu}}University of Trieste, I-34127 Trieste, Italy}
\affiliation{University of Tsukuba, Tsukuba, Ibaraki 305, Japan}
\affiliation{Tufts University, Medford, Massachusetts 02155, USA}
\affiliation{Waseda University, Tokyo 169, Japan}
\affiliation{Wayne State University, Detroit, Michigan 48201, USA}
\affiliation{University of Wisconsin-Madison, Madison, Wisconsin 53706, USA}
\affiliation{Yale University, New Haven, Connecticut 06520, USA}

\author{T.~Aaltonen}
\affiliation{Division of High Energy Physics, Department of Physics, University of Helsinki, FIN-00014, Helsinki, Finland; Helsinki Institute of Physics, FIN-00014, Helsinki, Finland}
\author{M.G.~Albrow}
\affiliation{Fermi National Accelerator Laboratory, Batavia, Illinois 60510, USA}
\author{S.~Amerio\ensuremath{^{ll}}}
\affiliation{Istituto Nazionale di Fisica Nucleare, Sezione di Padova, \ensuremath{^{ll}}University of Padova, I-35131 Padova, Italy}
\author{D.~Amidei}
\affiliation{University of Michigan, Ann Arbor, Michigan 48109, USA}
\author{A.~Anastassov\ensuremath{^{w}}}
\affiliation{Fermi National Accelerator Laboratory, Batavia, Illinois 60510, USA}
\author{A.~Annovi}
\affiliation{Laboratori Nazionali di Frascati, Istituto Nazionale di Fisica Nucleare, I-00044 Frascati, Italy}
\author{J.~Antos}
\affiliation{Comenius University, 842 48 Bratislava, Slovakia; Institute of Experimental Physics, 040 01 Kosice, Slovakia}
\author{G.~Apollinari}
\affiliation{Fermi National Accelerator Laboratory, Batavia, Illinois 60510, USA}
\author{J.A.~Appel}
\affiliation{Fermi National Accelerator Laboratory, Batavia, Illinois 60510, USA}
\author{T.~Arisawa}
\affiliation{Waseda University, Tokyo 169, Japan}
% ...

\collaboration{CDF Collaboration}
\altaffiliation[With visitors from]{
\ensuremath{^{a}}University of British Columbia, Vancouver, BC V6T 1Z1, Canada,
\ensuremath{^{b}}Istituto Nazionale di Fisica Nucleare, Sezione di Cagliari, 09042 Monserrato (Cagliari), Italy,
\ensuremath{^{c}}University of California Irvine, Irvine, CA 92697, USA,
\ensuremath{^{d}}Institute of Physics, Academy of Sciences of the Czech Republic, 182~21, Czech Republic,
\ensuremath{^{e}}CERN, CH-1211 Geneva, Switzerland,
\ensuremath{^{f}}Cornell University, Ithaca, NY 14853, USA,
\ensuremath{^{g}}University of Cyprus, Nicosia CY-1678, Cyprus,
\ensuremath{^{h}}Office of Science, U.S. Department of Energy, Washington, DC 20585, USA,
\ensuremath{^{i}}University College Dublin, Dublin 4, Ireland,
\ensuremath{^{j}}ETH, 8092 Z\"{u}rich, Switzerland,
\ensuremath{^{k}}University of Fukui, Fukui City, Fukui Prefecture, Japan 910-0017,
\ensuremath{^{l}}Universidad Iberoamericana, Lomas de Santa Fe, M\'{e}xico, C.P. 01219, Distrito Federal,
\ensuremath{^{m}}University of Iowa, Iowa City, IA 52242, USA,
\ensuremath{^{n}}Kinki University, Higashi-Osaka City, Japan 577-8502,
\ensuremath{^{o}}Kansas State University, Manhattan, KS 66506, USA,
\ensuremath{^{p}}Brookhaven National Laboratory, Upton, NY 11973, USA,
\ensuremath{^{q}}Istituto Nazionale di Fisica Nucleare, Sezione di Lecce, Via Arnesano, I-73100 Lecce, Italy,
\ensuremath{^{r}}Queen Mary, University of London, London, E1 4NS, United Kingdom,
\ensuremath{^{s}}University of Melbourne, Victoria 3010, Australia,
\ensuremath{^{t}}Muons, Inc., Batavia, IL 60510, USA,
\ensuremath{^{u}}Nagasaki Institute of Applied Science, Nagasaki 851-0193, Japan,
\ensuremath{^{v}}National Research Nuclear University, Moscow 115409, Russia,
\ensuremath{^{w}}Northwestern University, Evanston, IL 60208, USA,
\ensuremath{^{x}}University of Notre Dame, Notre Dame, IN 46556, USA,
\ensuremath{^{y}}Universidad de Oviedo, E-33007 Oviedo, Spain,
\ensuremath{^{z}}CNRS-IN2P3, Paris, F-75205 France,
\ensuremath{^{aa}}Universidad Tecnica Federico Santa Maria, 110v Valparaiso, Chile,
\ensuremath{^{bb}}Sejong University, Seoul 143-747, Korea,
\ensuremath{^{cc}}The University of Jordan, Amman 11942, Jordan,
\ensuremath{^{dd}}Universite catholique de Louvain, 1348 Louvain-La-Neuve, Belgium,
\ensuremath{^{ee}}University of Z\"{u}rich, 8006 Z\"{u}rich, Switzerland,
\ensuremath{^{ff}}Massachusetts General Hospital, Boston, MA 02114 USA,
\ensuremath{^{gg}}Harvard Medical School, Boston, MA 02114 USA,
\ensuremath{^{hh}}Hampton University, Hampton, VA 23668, USA,
\ensuremath{^{ii}}Los Alamos National Laboratory, Los Alamos, NM 87544, USA,
\ensuremath{^{jj}}Universit\`{a} degli Studi di Napoli Federico II, I-80138 Napoli, Italy
}
\noaffiliation


\date{\today}

\begin{abstract}
A measurement of the inclusive production  cross section of isolated prompt photons in proton-antiproton collisions at center-of-mass energy $\sqrt{s}$~=~1.96~TeV is presented. The results are obtained using the full Run II data sample collected with the Collider Detector at the Fermilab Tevatron, which corresponds to an integrated luminosity of 9.5~fb$^{-1}$. The cross section is measured as a function of photon transverse energy, $E_T^{\gamma}$, in the range 30~$ <  E_T^{\gamma} <$~500~GeV and in the pseudorapidity region $|\eta^{\gamma}|<$~1.0. The results are compared with predictions from parton-shower Monte Carlo models at leading order in quantum chromodynamics (QCD) and from next-to-leading order perturbative QCD calculations. The latter show good agreement with the measured cross section.
\end{abstract}

\pacs{12.38.Qk, 13.85.Qk}
\maketitle

\section{Introduction}
\label{sec:introduction}
The measurement of the cross section for the production of inclusive prompt photon ($\gamma$) in  proton-antiproton ($p\bar{p}$) collisions is an important test for perturbative quantum chromodynamics (pQCD), probing the parton distribution functions (PDFs),  and the parton-to-photon fragmentation functions (FFs)~\cite{owens, vogel, bourhis}. In addition, prompt-photon production is a major background for many other  standard model (SM) processes such as Higgs-boson decays into photon pairs ($ H \to \gamma \gamma$) and  in searches for non-SM physics with final states containing photons~\cite{binoth, randall, dine}. \\
\indent The term ``prompt'' identifies photons that are produced directly in the hard interaction and do not arise from hadron decays. 
In  $p\bar{p}$ collisions, events with prompt photons with transverse energy $E^{\gamma}_T$~\footnote{A cylindrical coordinate system $(r, \phi, z)$ is used with origin at the geometric center of the detector; $r$ is the radius from the nominal beam line, $\phi$ is the azimuthal angle, and the $+z$ axis points along the incident proton beam direction. 
The polar angle $\theta$ with respect to the proton beam is used to define the pseudorapidity $\eta = - \ln ( \tan (\theta/2))$.  Transverse energy and transverse momentum are defined as $E_T = E \sin(\theta)$  and $p_T = p \sin(\theta)$, respectively. The missing transverse energy is given by $\not{\hspace{-0.5ex}\vec{E}}_\textrm{T} = - \sum_i E_T^i \hat{n}_i $ where $i$ is the calorimeter tower number and $\hat{n}_i$ is a unit vector perpendicular to the beam axis and pointing at the $i$th calorimeter tower.} 
smaller than approximately~100~GeV are produced predominantly via quark-gluon Compton scattering $ qg \to q\gamma $, while at higher energies the quark-antiquark annihilation process $ q\bar{q} \to g\gamma $  plays a dominant role. In addition, prompt photons are produced through the collinear fragmentation of a final-state parton into a photon, e.g., $ qq \to qq \to \gamma X $, where $X$ can be any set of final state particles.\\ 
\indent The first measurement of the prompt-photon production cross section in hadron collisions came from the CERN Intersecting Storage Rings $pp$ collider, followed by measurements at the $Sp\bar{p}S$ collider~\cite{ISR_a, ISR, UA1, UA6}. More recent prompt-photon measurements have been performed at the Fermilab Tevatron Collider by the CDF and D0 Collaborations using $p\bar{p}$ collisions collected at a center-of-mass energy $\sqrt{s}$~=~1.8~TeV and 1.96~TeV~\cite{PhysRevD.65.112003, deluca, d0} and at the CERN Large Hadron Collider by the ATLAS and CMS Collaborations using $pp$ collisions at  $\sqrt{s}$~=~7~TeV~\cite{PhysRevD.83.052005, Aad2011150, PhysRevD.89.052004, PhysRevLett.106.082001, PhysRevD.84.052011}, 8~TeV~\cite{hep.ex.atlas}, and 13~TeV~\cite{2hep.ex.atlas}.\\ 
\indent This  article presents a measurement of the inclusive cross section for isolated prompt photons  over the range  30~$ <  E_T^{\gamma} <$~500~GeV, based on the full data set collected by the Collider Detector (CDF) during Run II (2001--2011) of the Fermilab Tevatron collider and corresponding to an integrated luminosity of 9.5~fb$^{-1}$~\footnote{The data set for this measurement corresponds to an integrated luminosity four times larger than that of the previously published CDF measurement~\cite{deluca}}.

\section{The CDF II detector}
\label{sec:cdf} 
The CDF~II detector~\cite{cdf} is a general-purpose spectrometer at the Fermilab Tevatron collider. It has a cylindrical geometry with approximate forward-backward and azimuthal symmetry. 
It includes a charged-particle tracking system consisting of silicon microstrip detectors and a cylindrical open-cell drift chamber, designed to  measure charged-particle trajectories (tracks) and momenta. The tracking system is contained within a 1.4~T axial magnetic field. It is surrounded by  electromagnetic~(EM) and hadronic calorimeters segmented in projective towers and  used to identify and measure the energy and position of photons, electrons, hadrons, and clusters of particles (jets). The central calorimeters cover the region $|\eta| < 1.1$, and have electromagnetic transverse-energy resolution of $\sigma (E_T) / E_T = 13.5\% / \sqrt{E_T (\rm{GeV})} \oplus  1.5\%$ and a tower segmentation of $\Delta \eta \times \Delta \phi \approx 0.1 \times$\ang{15} in pseudorapidity-angular space~\cite{Note1}. 
At a depth corresponding approximately to the maximum energy density in the development of a typical EM shower, the EM calorimeters contain detectors that measure  the transverse shower profile. 
The electromagnetic compartments of the calorimeter are equipped with a timing system measuring the arrival time of particles that deposit energy in each tower~\cite{em2}.
Drift chambers and scintillation counters located outside the calorimeters identify muons. 


\section{Data and simulated samples}
\label{sec:dataset}
\subsection{Event selection}
\label{sec:sel}
Photons are reconstructed using clusters of (up to three) adjacent towers above threshold in the central EM calorimeter~\cite{PhysRevD.82.052005}. The pseudorapidity  is restricted to the fiducial region $|\eta^{\gamma}|<\rm{1.0}$. The data are collected using a three-level online event-filtering system (trigger)~\cite{trigger} that selects events with at least one EM cluster consistent with a photon in the final state. Since there can be multiple collisions in the same bunch crossing, the event primary vertex ($p\bar{p}$~interaction point) is chosen to be the one that results in the best balance of the $p_T$ of the photon; the $z$ position of the reconstructed primary vertex is required to be within 60~cm of the center of the detector.
The photon transverse energy is corrected to account for nonuniformities in the calorimeter response, and calibrated using electrons from reconstructed $Z$-boson decays~\cite{calib}.  Photon candidates are required to satisfy $E_T^{\gamma}>$~30~GeV and to meet requirements on calorimeter isolation~\footnote{The calorimeter isolation is defined as the transverse energy deposits in the EM calorimeter in the isolation cone minus the transverse energy in the EM cluster of the photon. The isolation cone is defined to have a radius $R =\sqrt{(\Delta \eta)^2 + (\Delta \phi)^2 } =$~0.4. The track isolation is defined as the scalar sum of the transverse momenta of all tracks originating from the primary vertex of the event and lying within a cone of radius $R =$~0.4.}, on track isolation~\cite{Note3}, and on the ratio of the energy deposited in the hadronic calorimeter to the energy in the EM cluster~\cite{PhysRevD.82.052005}. 
If more than one prompt photon candidate is reconstructed ($\approx$~1\% of the photon events), that with the highest $E_T$ (leading photon) is chosen.\\
\indent Events with electrons from $Z$- and $W$-boson decays, which can be misidentified as photons, are removed from the sample by requiring 0 tracks, or at the most one soft track (track isolation $\leq$5~GeV), pointing to the EM cluster. This track is allowed to account for underlying event and pile-up energy around the cluster. The number of electrons coming from $W$-boson decays is  further reduced by requiring the missing transverse energy~\cite{Note1} of the event to be less than 80\% of the transverse energy of the  photon candidate. This requirement also reduces backgrounds arising from other  sources that lead to energy imbalance, such as muons from cosmic rays that may  emit bremsstrahlung radiation in the calorimeter,  and muons from beam-halo interactions with the beam pipe, which may in turn interact with the detector material producing photons.
Finally, the EM signal timing is required to be consistent with the $p\bar{p}$ collision time~\cite{PhysRevD.82.052005}. The residual backgrounds from $Z$- and $W$-boson decays, cosmic rays and beam halo,  are expected to be less than 1\% of the total sample.\\
\indent After applying all the selection criteria,  2.1$\times$10$^6$ events remain in the $\gamma+X$ data sample. The dominant remaining backgrounds are due to jets misidentified as photons.
\subsection{Simulated events}
\label{sec:mc}
Simulated events from the \textsc{pythia} Monte-Carlo (MC) generator~\cite{pythia} are used in the background estimation, and to evaluate the product of the detector acceptance ($A$) and the efficiency ($\epsilon$) for signal events. 
 Monte Carlo samples are generated with \textsc{pythia}~6.216, a parton-shower generator at leading-order (LO) in the strong-interaction coupling,  with the CTEQ5L~PDFs~\cite{pythia}. 
The \textsc{pythia} predictions include  2~$ \to $~2 matrix-element  subprocesses. Higher-order QCD corrections  are included by initial- and final-state parton showers.\\
\indent For the study of systematic uncertainties and for comparisons with the final results, events are also generated with the \textsc{sherpa}~1.4.1 MC generator ~\cite{sherpa} with CT10~PDFs~\cite{sherpa2}. 
The \textsc{sherpa} predictions  include all the tree-level matrix-element amplitudes with one photon and up to three partons. This calculation features a parton-jet matching procedure to avoid an overlap between the phase-space descriptions given by the fixed-order matrix-element subprocesses and the showering and hadronization in the multijet simulation.\\
\indent The \textsc{Tune A}~\cite{Field:2005sa, tune} underlying event~\footnote{
The underlying event is that part of the event final state that cannot be directly associated with the primary hard 2$\to$2 parton-parton scattering and consists of the beam remnants plus possible
contributions from initial- and final-state gluon radiation and additional parton-parton interactions.} model is used in the \textsc{pythia} calculation. Monte Carlo events are passed through a \textsc{geant}-based simulation of the detector~\cite{geant} and subjected to the same reconstruction and selection requirements as the data.

\section{Signal fraction}
\label{sec:pi0_bkg}
\begin{figure}[h]
  \begin{center}
   \includegraphics[width=0.475\textwidth]{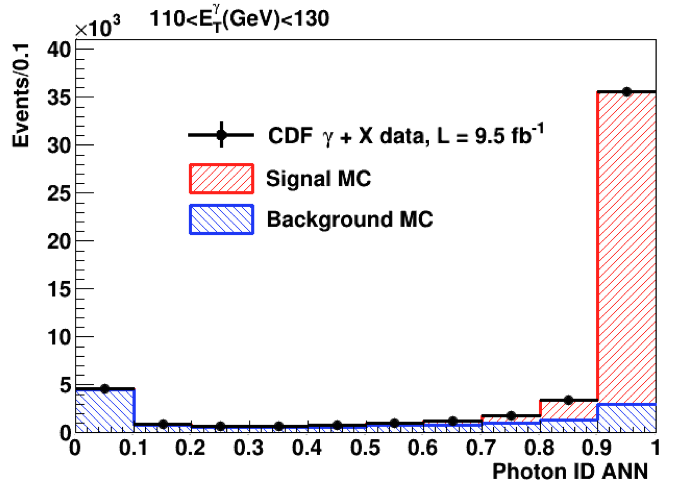}
    \caption{Observed ANN output distribution (points), the templates for signal and background processes, and the resulting fit of the templates to the data distribution, for events restricted to the photon transverse energy bin 110--130~GeV. The left-hatched histogram (blue in color) represents the background and the right-hatched histogram (red in color) represents the signal, normalized so that the scale of the sum of the templates equals the total number of data events.}
\label{fig:fit_bin08}
  \end{center}
\end{figure}

\begin{figure}[h]
  \begin{center}
    \includegraphics[width=0.475\textwidth]{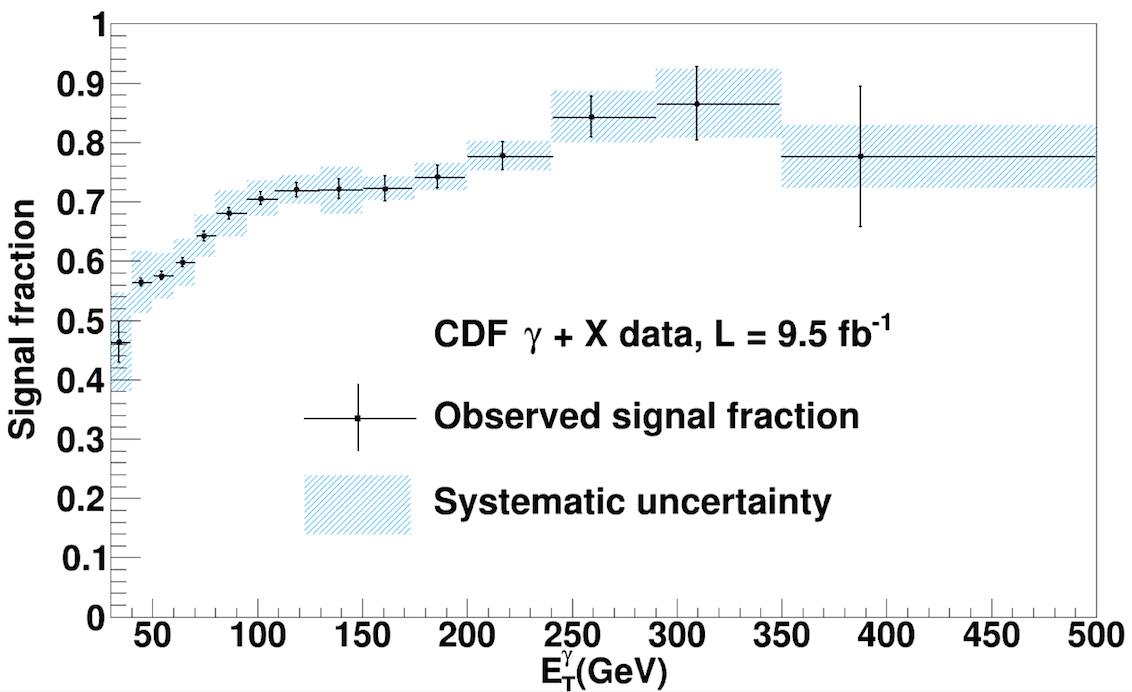}
    \caption{Signal fraction as a function of leading-photon $E_T$. 
     The points are plotted at the average $E_T^{\gamma}$, of the data within each bin. 
    The error bars represent the statistical uncertainty, while bands represent the total systematic uncertainty.}
\label{fig:sig_frac}
  \end{center}
\end{figure}

 After the event selection,  the remaining background comes from the decays of hadrons (such as $\pi^0 \to  \gamma \gamma$); they cannot be rejected on an event-by-event basis, so a statistical background-subtraction technique is used to measure the signal cross section. To evaluate the signal fraction, an artificial neural network (ANN) is defined using as input the shower-shape, transverse profile, and isolation variables~\cite{karen}. 
The  inclusive-photon simulation is matched to data by applying the same corrections as derived in Refs.~\cite{deluca, tin}. Further, MC events are reweighted to the observed instantaneous luminosity profile to account for luminosity-dependent effects. 
The expected ANN output distributions (``templates'') for  signal and background samples are constructed using  \textsc{pythia} inclusive-photon and dijet MC predictions, respectively. These templates are validated using the $Z\to e^{+}e^{-}$  and dijet data samples~\cite{tin}. 
To estimate the prompt-photon rate,  the ANN output distribution observed in data is fit to a linear combination of signal and background  ANN templates, 
using a binned maximum-likelihood method that accounts for uncertainties on both data and templates~\footnote{As implemented in \textsc{TFractionFitter} which is a class of the CERN \textsc{Root} analysis software~\cite{root, barlow}. The fit fractions are provided with an error estimate which takes into account both data and Monte Carlo statistical uncertainties. \textsc{TFractionFitter} errors are corrected based on pseudoexperiments~\cite{aluca}.}. 
A fit is performed in each $E_T^{\gamma}$ bin, yielding prompt-photon fractions in the  $E_T^{\gamma}$ range from 30~GeV up to 500~GeV, as shown in Fig.~\ref{fig:fit_bin08} for an example $E_T^{\gamma}$  bin.
 Figure~\ref{fig:sig_frac} shows the resulting signal fraction (photon purity) as a function of $E_T^{\gamma}$.
\\
\indent The systematic uncertainty on the signal fraction  is estimated by varying the fit configurations and  the values of the ANN input variables within their uncertainties. The dominant uncertainty on the  shape of the ANN templates  originates from the modeling of calorimeter isolation energy. The overall systematic uncertainty on the signal fraction is estimated to be 8\% at low $E_T^{\gamma}$,  6\% at high $E_T^{\gamma}$,  and 3\% on average for the intermediate  $E_T^{\gamma}$ range 40~$<E_T^{\gamma}<$~300~GeV.


\section{Cross section measurement}
\label{sec:xsec}
The differential cross section for the production of isolated prompt photons in a given phase-space bin is calculated as  $ d^2 \sigma  / ( d E^{\gamma}_T d \eta^{\gamma} ) =  (N  f^{\gamma}) /( \Delta  E^{\gamma}_T \Delta \eta^{\gamma} \mathcal{L} A \times \epsilon) \label{eq_xsec}$, where $N$ is the number of data events in a given $E_T^{\gamma}$ bin after applying the full selection, $f^{\gamma}$ is the signal fraction, $\Delta E_T^{\gamma}$  is the width of the $E_T^{\gamma}$ bin,  $ \mathcal{L} $ is the integrated luminosity, and $A\times\epsilon$ is a correction factor. Since the cross section is measured for $| \eta^{\gamma} |<1.0$, $\Delta \eta^{\gamma}$ is set to 2.0.\\
The factor $A\times\epsilon$ combines corrections for acceptance,  resolution effects and efficiencies for selecting and reconstructing the photon to infer the results at the particle level (\textit{i.e.}, generator level). 
The correction is computed from the bin-by-bin fraction of simulated particle-level prompt photons in the reconstructed signal events, as determined by the \textsc{pythia} MC calculation. 
The numerator is obtained by applying the same requirements to the \textsc{pythia}-simulated events as those applied to data. The denominator  is obtained by selecting  generated particles~\footnote{Generated particles are the stable particles, i.e., particles with a lifetime of at least 10~ps in events from MC generators, without any simulation of the interaction of these particles with the detector or any additional proton-antiproton interactions.} in the fiducial region, with $E_T^{\gamma}>$~30~GeV and the same energy isolation requirement as in the data. The photon efficiency is calibrated by comparing the selection efficiencies for $Z \to e^+ e^-$ events in data and in simulation~\cite{tin}.\\
\indent The largest sources of systematic uncertainty arise from the photon energy scale at high $E_T$ ($\approx$~6\%) and  from the $A\times\epsilon$  factor ($\approx$~8\%). The latter is determined by a comparison of results from the \textsc{pythia} and \textsc{sherpa} MC calculations.
\section{Theoretical Predictions}
\label{sec:theory}
The predicted prompt-photon production cross section is calculated using the fixed-order next-to-leading-order (NLO) program \textsc{mcfm}~6.8 including nonperturbative fragmentation at LO~\cite{mcfm}.
The calculation uses the MRST2008 NLO PDFs and the GdRG LO FFs~\cite{mcfm_gdrg}.
The \textsc{mcfm} prediction is a parton-level calculation that does not include a model for the underlying event energy. This prediction is corrected for the nonperturbative effects of parton-to-hadron fragmentation and for underlying event energy. 
A correction factor $C_{\rm{UE}} = 0.91 \pm 0.03$ is defined as the overall ratio of the  cross section obtained using the \textsc{pythia}  MC generator, with and without modeling of both multiple-parton interactions and hadronization~\cite{deluca}. 

The nominal renormalization ($\mu_R$), factorization ($\mu_F$), and fragmentation ($\mu_f$) scales are set to the photon transverse energy ($\mu_R = \mu_F = \mu_f = E_T^{\gamma}$). 
The scale uncertainty is evaluated by varying the three scales simultaneously between the extreme values $E_T^{\gamma} / 2$ and $2 E_T^{\gamma}$. \\
\indent In addition to comparison with the perturbative-QCD prediction above, we also compare the measured cross section to predictions from the \textsc{pythia} and \textsc{sherpa} MC generators. Both are calculated at the particle level, meaning that the photon isolation energy is estimated using generated hadrons and the selection criteria are applied to the hadron jets and  are directly comparable to our measurement.

\section{Results}
\label{sec:results}
The differential cross section results for the production of isolated prompt photons are listed in Table~\ref{tab:table1}, together with statistical and systematic uncertainties. 
\begin{table}[h]
\caption{\label{tab:table1}
Measured cross section for the production of prompt isolated photons within the pseudorapidity region $|\eta^{\gamma}|<1.0$, in bins of $E_T^{\gamma}$. $ \langle E_T^{\gamma} \rangle$, the average $E_T^{\gamma}$ within each bin, is listed for illustration of the steeply falling spectral shape. The measured-cross-section uncertainties given are statistical only. The column $\delta \sigma_{syst}$ gives the systematic uncertainties. The additional 6\% luminosity uncertainty is not included in the table.
}
\begin{ruledtabular}
\begin{tabular}{cclc}
 $E_T^{\gamma}$& $\langle E_T^{\gamma} \rangle $& $d^2 \sigma / (d E_T^{\gamma} d \eta^{\gamma}) $  & $\delta \sigma_{\rm{syst}}$\\
  (GeV)    & (GeV) & (pb/GeV)                                                                    &  (\%)\\
\hline
30--40     &   34.1 &  (5.49  $\rm{\pm}$ 0.41)$\rm{\times 10^{1} }$    &  23.3 \\
40--50     &   44.3 &  (1.72  $\rm{\pm}$ 0.23)$\rm{\times 10^{1} }$    &  17.2 \\
50--60     &   54.3 &  (6.72 $\rm{\pm}$  0.11)$\rm{\times 10^{0} }$    &  14.9 \\
60--70     &   64.4 &  (2.95  $\rm{\pm}$ 0.04)$\rm{\times 10^{0} }$    &  14.6 \\
70--80     &   74.5 &  (1.45  $\rm{\pm}$ 0.02)$\rm{\times 10^{0} }$    &  13.7 \\
80--90     &   86.5 &  (6.87  $\rm{\pm}$ 0.10)$\rm{\times 10^{-1} }$  &  13.2 \\
90--110   & 101.7 & (3.03  $\rm{\pm}$ 0.05)$\rm{\times 10^{-1} }$  &  12.8 \\
110--130 & 118.7 & (1.32  $\rm{\pm}$ 0.03)$\rm{\times 10^{-1} }$  &  12.7 \\
130--150 & 138.8 & (5.65  $\rm{\pm}$ 0.15)$\rm{\times 10^{-2} }$  &  13.1 \\
150--175 & 160.9 & (2.37  $\rm{\pm}$ 0.08)$\rm{\times 10^{-2} }$  &  12.6 \\
175--200 & 185.9 & (1.03  $\rm{\pm}$ 0.03)$\rm{\times 10^{-2} }$  &  12.4 \\
200--240 & 216.8 & (4.01  $\rm{\pm}$ 0.12)$\rm{\times 10^{-3} }$  &  13.2 \\
240--290 & 259.2 & (1.16  $\rm{\pm}$ 0.05)$\rm{\times 10^{-3} }$  & 14.1 \\
290--350 & 309.4 & (3.08  $\rm{\pm}$ 0.23)$\rm{\times 10^{-4} }$  &  15.1 \\
350--500 & 387.6 & (1.83  $\rm{\pm}$ 0.29)$\rm{\times 10^{-5} }$  &  16.1 \\

\end{tabular}
\end{ruledtabular}
\end{table}
The systematic uncertainties on the differential cross section are determined by propagating  the sources of uncertainty considered for  $f^{\gamma}$ and $ A \times \epsilon$.
At low $E_T^{\gamma}$ the total systematic uncertainty is dominated by the uncertainties in the ANN-template modeling, while the dependence of the $ A \times \epsilon$ factors on the event generator  gives the dominant contribution ($\approx$~10\%) to the uncertainty at intermediate and high $E_T^{\gamma}$. 
Finally, there is an additional 6\% uncertainty on the integrated luminosity~\cite{lumi}.

\begin{figure}[h]
  \begin{center}
\includegraphics[width=0.475\textwidth]{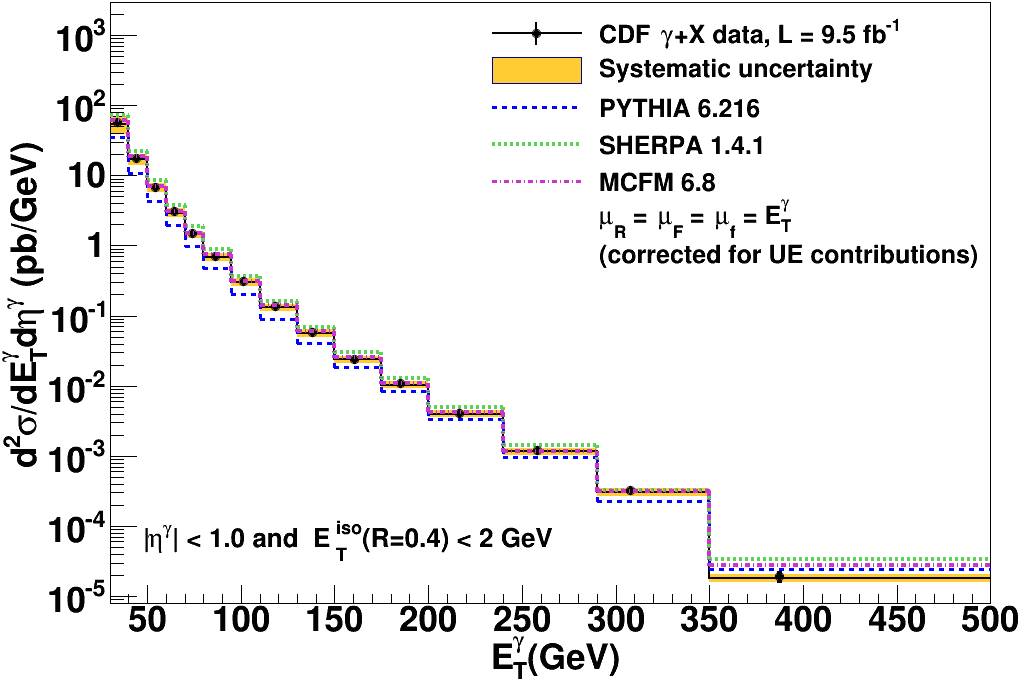}  
    \caption{Measured $\gamma + X$ cross section as a function of leading photon  transverse energy. Data (markers) are centered at the average $E_T^{\gamma}$ of each bin for illustration of the steeply falling spectral shape. Data are compared with the \textsc{pythia}, \textsc{sherpa} and \textsc{mcfm} predictions (dashed lines). The vertical error bars show the statistical uncertainties, while the shaded areas show the systematic uncertainties. The 6\% luminosity uncertainty on the data is not included.
A correction $C_{\rm{UE}}$  to account for parton-to-hadron fragmentation and underlying event effects is applied to the \textsc{mcfm} theoretical predictions, as explained in the text.}
\label{fig:XSec}
  \end{center}
\end{figure}


\begin{figure}[h]
  \begin{center}
    \includegraphics[width=0.475\textwidth]{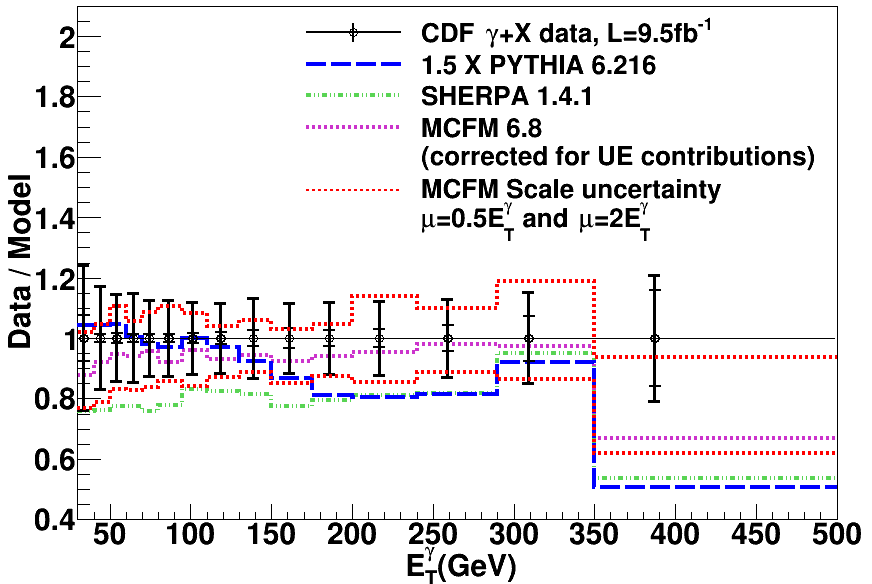}  
    \caption{Data points centered at 1.0 and data-to-theory ratio (dashed lines) of the inclusive prompt-photon differential cross section as a function of the photon transverse energy, in the pseudorapidity region $|\eta^\gamma| <$~1.0. The inner error bars on the data points show statistical uncertainties. The full error bars show statistical and systematic uncertainties added in quadrature. The 6\% luminosity uncertainty on the data is not included. 
The LO \textsc{pythia} prediction is multiplied by a factor 1.5.}
\label{fig:ratioXSec}

  \end{center}
\end{figure}
These results are compared with the theoretical predictions in Fig.~\ref{fig:XSec}.
The ratio of the measured cross section over the predicted ones is shown in Fig.~\ref{fig:ratioXSec}. The full error bars on the data points represent statistical and systematic uncertainties summed in quadrature. The inner error bars show statistical uncertainties only. The NLO predictions are shown with their theoretical uncertainties arising from the choice of factorization, renormalization, and fragmentation scales.\\
The NLO calculations agree with the data up to the highest $E_T^{\gamma}$-values considered. Observed cross sections are moderately larger than the central values for the NLO calculation for low $E_T^{\gamma}$, but agree within the theoretical uncertainty of the NLO calculation.\\
The \textsc{pythia} and \textsc{sherpa} predictions are also shown in Figs.~\ref{fig:XSec} and \ref{fig:ratioXSec}. The shape of the measured-cross-section distribution is well described by both models. The \textsc{pythia} prediction underestimates the observed cross section by more than a factor of 1.5 across the whole  $E_T^{\gamma}$ range. This is possibly due to the lack of higher-order terms in the \textsc{pythia} photon+jet matrix-elements. The  \textsc{sherpa} calculation is approximately 1.1 to 1.2 times larger than the observed cross section, nearly uniformly  across the $E_T^{\gamma}$ range. This calculation includes up to three jet emissions associated with the observed photon, but it is missing virtual corrections in the matrix elements of the subprocesses, which could possibly explain the discrepancy with data. Other possible reasons are related to nonperturbative QCD processes, such as mistuned fragmentation subprocesses leading to excessive rates of photon production through fragmentation.
\section{Conclusions}
\label{sec:concl}
A measurement of the differential cross section for the inclusive production of isolated prompt photons in $p\bar{p}$ collisions at $\sqrt{s}$~=~1.96~TeV is presented using the full data set collected with the CDF~II detector at the Tevatron. The cross section is measured as a function of photon transverse energy $E_T^{\gamma}$ in the central pseudorapidity region $|\eta^{\gamma}|<1.0$. The measurement spans the $E_T^{\gamma}$ kinematic range from 30~GeV to 500~GeV, thus extending the reach by 100~GeV from the previous CDF measurement~\cite{deluca}.
Comparisons of our measurement to three theoretical predictions are discussed. 
Both \textsc{pythia} and \textsc{sherpa} predictions correctly describe the shape of the differential cross section. The \textsc{pythia} generator predicts a smaller cross section compared to the data and the \textsc{sherpa} prediction.
The data are in  good agreement with the fixed-order NLO \textsc{mcfm} calculation.\\

 

\section*{Acknowledgements}
\label{sec:acknow}
We thank the Fermilab staff and the technical staffs of the
participating institutions for their vital contributions. This work
was supported by the U.S. Department of Energy and National Science
Foundation; the Italian Istituto Nazionale di Fisica Nucleare; the
Ministry of Education, Culture, Sports, Science and Technology of
Japan; the Natural Sciences and Engineering Research Council of
Canada; the National Science Council of the Republic of China; the
Swiss National Science Foundation; the A.P. Sloan Foundation; the
Bundesministerium f\"ur Bildung und Forschung, Germany; the Korean
World Class University Program, the National Research Foundation of
Korea; the Science and Technology Facilities Council and the Royal
Society, United Kingdom; the Russian Foundation for Basic Research;
the Ministerio de Ciencia e Innovaci\'{o}n, and Programa
Consolider-Ingenio 2010, Spain; the Slovak R\&D Agency; the Academy
of Finland; the Australian Research Council (ARC); and the EU community
Marie Curie Fellowship Contract No. 302103.


\bibliography{apssamp_InclPhotonXsec}

\end{document}